\documentclass[11pt]{elsart}
\usepackage[dvips]{graphicx}
\usepackage{amsmath}
\usepackage{amssymb}

\begin{document}
\begin{frontmatter}

\title{The study of dynamic singularities  of seismic signals by the
 generalized Langevin equation}

\author[label1]{Renat Yulmetyev}
\ead{renat.yulmetyev@mail.ru}
\author[label1]{Ramil Khusnutdinoff}
\ead{khrm@mail.ru}
\author[label3]{Timur Tezel}
\author[label3]{Yildiz Iravul}
\author[label3]{Bekir Tuzel}
\author[label4]{Peter H\"anggi}
\address[label1]{Department of Physics, Kazan State
University, Kremlevskaya Street  18, 420008 Kazan, Russia}
\address[label3]{General Directorate of Disaster Affairs Earthquake
Research Department, Eskisehir yolu 10.km Lodumlu/ANKARA, Turkey}
\address[label4]{Department of Physics, University of Augsburg,
Universit\"atsstrasse 1, D-86135 Augsburg, Germany}

\begin{abstract}
Analytically and quantitatively we reveal that the GLE equation,
based on a memory function approach, in which memory functions and
information measures of statistical memory play fundamental role
in determining the thin details of the stochastic behavior of
seismic systems, naturally conduces to a description of seismic
phenomena in  terms of  strong and weak memory. Due to a
discreteness of  seismic signals we use a finite - discrete form
of GLE.  Here we studied some cases of seismic activities of Earth
ground motion  in Turkey with consideration of complexity,
nonergodicity and fractality  of seismic signals.
\\

\end{abstract}

\date{\today}
\end{frontmatter}


\section{Introduction}
Specific stochastic dynamics occur in a large variety of systems,
such as supercooled liquids, seismic systems, human brain,
finance, meteorology and granular matter. These systems are
characterized by an extremely rapid increase or a slowdown of
relaxation times and by a non-exponential decay of time-dependent
correlation functions \cite{Abe,Corral_2006}.

The canonical theoretical framework for stochastic dynamics of
complex systems  is the time-dependent generalized Langevin
equation (GLE) \cite{Zwanzig,Mori,Yulmetyev_2000,Yulmetyev_2001,Yulmetyev_2005,Goetze,Das}.
It successfully describes the phenomenon of
statistical memory, whereby the relaxation time for order
parameter fluctuations scales as a power of the correlation
length. An obvious question to ask would be whether this framework
can be adapted to describe seismic phenomena. Analytically and
quantitatively we show that the GLE equation, based on a memory
function approach, where the memory functions and information
measures of statistical memory play fundamental role in
determining the thin details of the stochastic behavior of seismic
systems, naturally leads to a description of seismic phenomena in
a terms of a strong and weak memory. Due the discreteness of a
seismic signals we use a finite - discrete form of GLE.  Here we
study some cases of seismic activities of Earth ground motion in
last years in Turkey with consideration of complexity,
irregularity and metastability of seismic signals.

\section{Some extraction from the theory of discrete stochastic processes}

The GLE analytical model was originally proposed for displaying
stochastic  behavior of  signals in complex systems
\cite{Zwanzig,Mori,Yulmetyev_2000,Yulmetyev_2001,Yulmetyev_2005}, one of
which identifies memory effects  with diverse memory time scales
in manyfold of signal's correlation  so that the arbitrary seismic
state is characterized by the set of  memory time length scale in
a system.

Here we consider  data of seismic signals recording as a time
series $\xi$:
\begin{equation}
\xi=\{ \xi_0, \xi_1, \xi_2,\ldots, \xi_{N-1}\}=\{ \xi(0),
\xi(\tau), \xi(2\tau),\ldots, \xi([N-1]\tau)\}.
\end{equation}
Here  $\tau$ is a discretization time of  seismic signals, $N$ is
a total number of signals. A set of fluctuations $\delta\xi$ is an
initial dynamic variable $W_0$ :
\begin{equation}
W_0=\{ \delta \xi_0, \delta \xi_1, \delta \xi_2,\ldots,
 \delta \xi_{N-1}\}, ~~~~~\delta \xi_j=\xi_j-\langle \xi\rangle, ~~~~~ \langle \xi\rangle
=\frac{1}{N}\sum_{j=0}^{N-1}\xi_j.
\end{equation}
The Gram-Schmidt orthogonalization procedure
\begin{equation}
\langle W_n,W_m \rangle=\delta_{n,m}\langle |W_n|^2 \rangle
\end{equation}
leads to the set of the following orthogonal dynamic variables:
\begin{equation}
\left \{
\begin{array}{l}
W_0=\delta \xi,\\
W_1=\mathcal{L}W_0=\displaystyle{\frac{d}{dt}\delta\xi}, \\
W_2=\mathcal{L}W_1-\Lambda_1W_0, \\
~~~~~~~~~~~~~~~~~~~~~~~~~~\ldots, \\
W_{n+1}=\mathcal{L}W_n-\Lambda_nW_{n-1},~~~~~n \geq 1,
\end{array}
\right.
\end{equation}
where $\mathcal{L} = (\Delta - 1)/ \tau$  is the Liouville's
quasioperator and $\Lambda_n$ is the relaxation parameter of the
n$th$ order (where $ \Delta  $ is the shift operator $ \Delta
x_{j} = x_{j + 1} $  and $\tau $ is the discretization time).

Within the  framework of  statistical theory and Zwanzig-Mori's
theoretical-functional procedure of projection operators one can
receive  following recurrent relation as a finite-difference
kinetic equation:

\begin{equation}
\Delta M_n(t) =\tau\lambda_{n+1}
M_n(t)-\tau^2\Lambda_{n+1}\sum_{j=0}^{m-1}M_{n+1}(t-j\tau)M_n(j\tau),
~~ n=0,1,2,\ldots.
\end{equation}
Here we introduce a Liouville's quasioperator eigenvalue
$\lambda_{n+1}$, a relaxation parameter $\Lambda_{n+1}$ and a
memory function $M_n(t)$ of the n$th$  order respectively

\begin{equation}
\lambda_n=\frac{\langle W_{n-1}\mathcal{L}W_{n-1}\rangle}{\langle
|\langle W_{n-1}|^2\rangle},~~~~~ \Lambda_n=\frac{\langle |
W_{n}|^2\rangle}{\langle | W_{n-1}|^2\rangle} ,~~~~~
M_n(t)=\frac{\langle  W_{n}(t)W_{n}\rangle}{\langle |
W_{n}|^2\rangle}.
\end{equation}

For analysis of relaxation time scales of underlying processes we
use the frequency-dependent statistical non-Markovity parameter
$\varepsilon_n (\omega)$:

\begin{equation}
\varepsilon_n (\omega) = \left\{ \frac {\mu _ {n-1} (\omega)} {
\mu_n (\omega)} \right\}^ {1 / 2}.
\end{equation}

Here $\mu_n(\omega)$ is a frequency power spectrum for the memory
function of the  n$th$  order:
\begin{equation}
\mu_n(\omega)=\left |\tau \sum_{j=0}^{N-1}M_n(j\tau)\cos
(j\tau\omega)\right |^2.
\end{equation}

Using of Eqns. (1) - (8) we can study all specific singularities
of the statistical  memory effects in an underlying system.
Non-Markovity parameter and its statistical spectrum were
introduced by Yulmetyev et al. in \cite{Yulmetyev}. It is worthy
of  mentioning that non-Markovian character of seismic data  was
discussed by Varotsos et al. \cite{Varotsos}. One of the first
proofs of non-Markovity of empirical random processes was given in
Refs. \cite{Fulinski}. Stochastic origins of the long-range
correlations of ionic current fluctuations in membrane channels
with non-Markovian behavior were studied in \cite{Weron}.

\begin{figure}
\begin{center}
\includegraphics[width=1.0\textwidth]{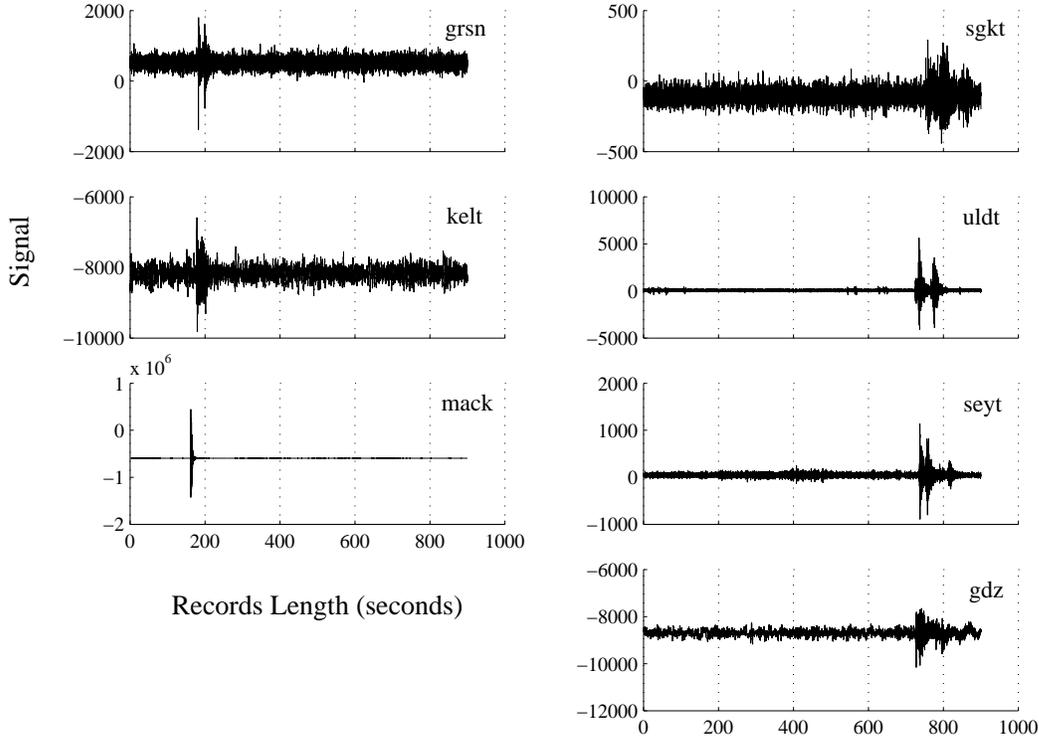}
\end{center}
\caption[kurzform]{\label{TS} An initial row data of seismic signals from 7
areas of seismic signals: $grsn$, $kelt$, $mack$, $sgkt$, $uldt$,
$seyt$, $gdz$. The discretization time is $\tau=0.02$ sec.}
\end{figure}

\section{An analysis of results}

\begin{figure}
\begin{center}
\includegraphics[width=1.0\textwidth]{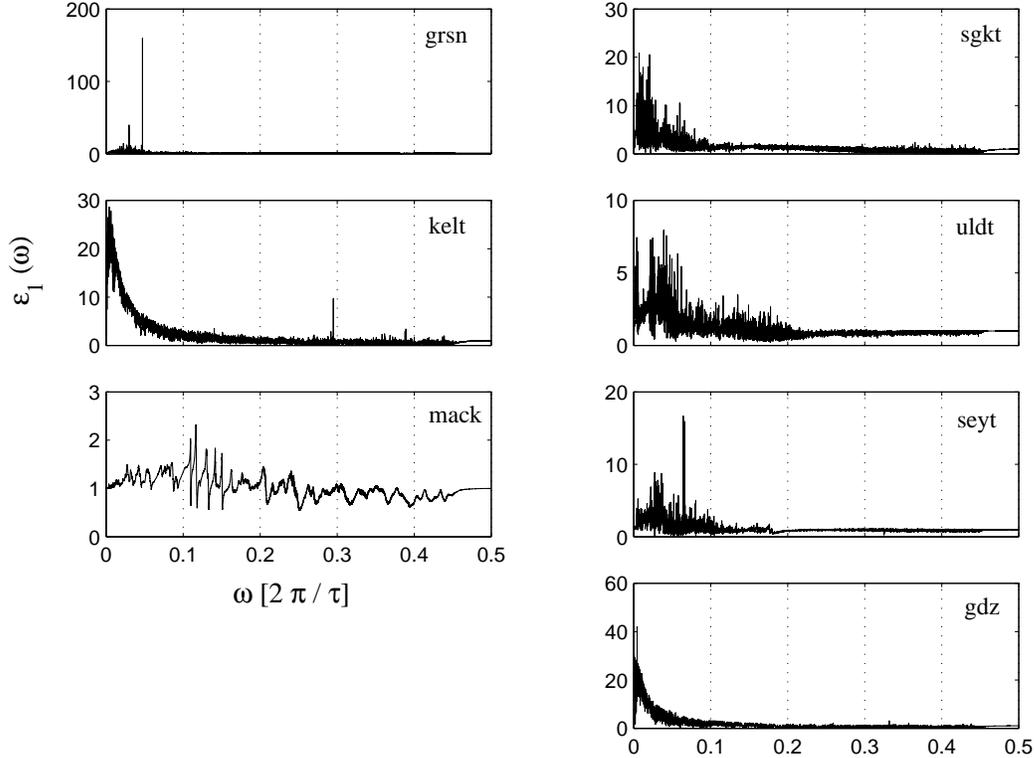}
\end{center}
\caption[kurzform]{\label{MP} The frequency dependence of the first point of
non-Markovity parameter  $\varepsilon_1(\omega)$ for the each
seismic origin: $grsn$, $kelt$, $mack$, $sgkt$, $uldt$, $seyt$,
$gdz$.}
\end{figure}

Fig. \ref{TS} presents the initial time series of seismic signals
for 7 seismic origins: $grsn$, $kelt$, $mack$, $sgkt$, $uldt$,
$seyt$, $gdz$.  Discretization time is $\tau=0.02$ sec. From the
Figures we can see that all time series have distinctive features.

Fig. \ref{MP} demonstrates the frequency dependence of the first
point of non-Markovity parameter for 7 seismic origins from Turkey
$\varepsilon_1(\omega)$: $grsn$, $kelt$, $mack$, $sgkt$, $uldt$,
$seyt$, $gdz$ . Since the nature of each seismic source  is
unknown to us, it would be interesting to establish its character.
It seems possible that the  signals can be distributed to 3
groups: \textbf{group A} ($kelt$, $gdz$), \textbf{group B}
($grsn$, $sgkt$, $uldt$, $seyt$) and \textbf{group C} ($mack$).

\textbf{Signals of Group  A} are characterized by the more
regular structure and smooth decay of the function
$\varepsilon_1(\omega)$.

\begin{figure}
\begin{center}
\includegraphics[width=0.8\textwidth]{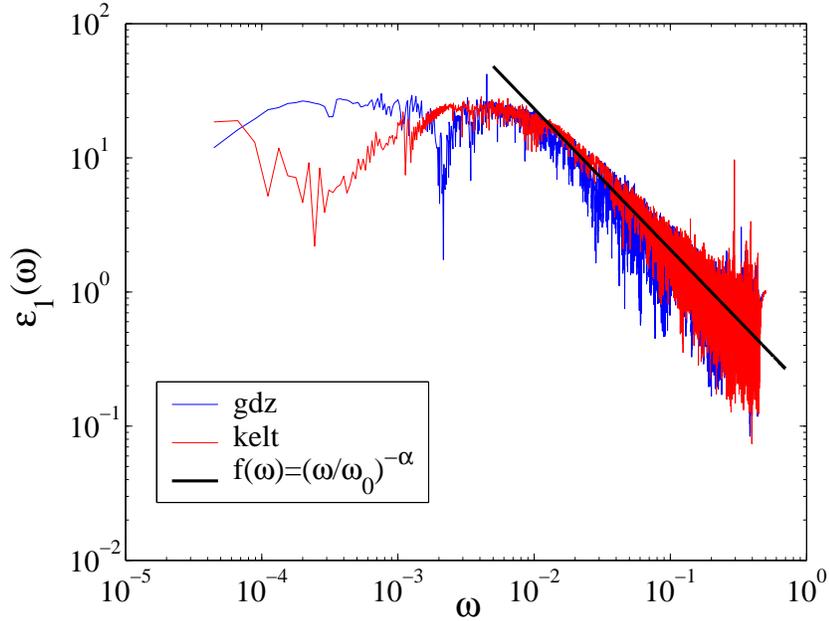}
\end{center}
\caption[kurzform]{\label{Pnm_gdz_kelt02} (Color online) The frequency dependence of
$\varepsilon_1(\omega)$ in double log-log scale for 2 seismic
origins: $gdz$ è $kelt$. The time discretization is $\tau=0.02$ sec.
The general power dependence of
$\varepsilon_{1}(\omega)=(\omega/\omega_0)^{-\alpha}$ is submitted
by a continuous line with parameters $\omega_0=0.2$,
$\alpha=1.05$.}
\end{figure}

There is a frequency dependence of non-Markovity parameter for
seismic origins: $gdz$ è $kelt$  in Fig. \ref{Pnm_gdz_kelt02}.
General power dependence
$\varepsilon_{1}(\omega)=(\omega/\omega_0)^{-\alpha}$ is submitted
by a continuous line with parameters $\omega_0=0.2$,
$\alpha=1.05$.

\textbf{Signals  of Group B} are characterized by the irregular
frequency structure and by the frequency bursts on the distinct
frequencies. Spectra have a noisy character.

\begin{figure}
\begin{center}
\includegraphics[width=0.8\textwidth]{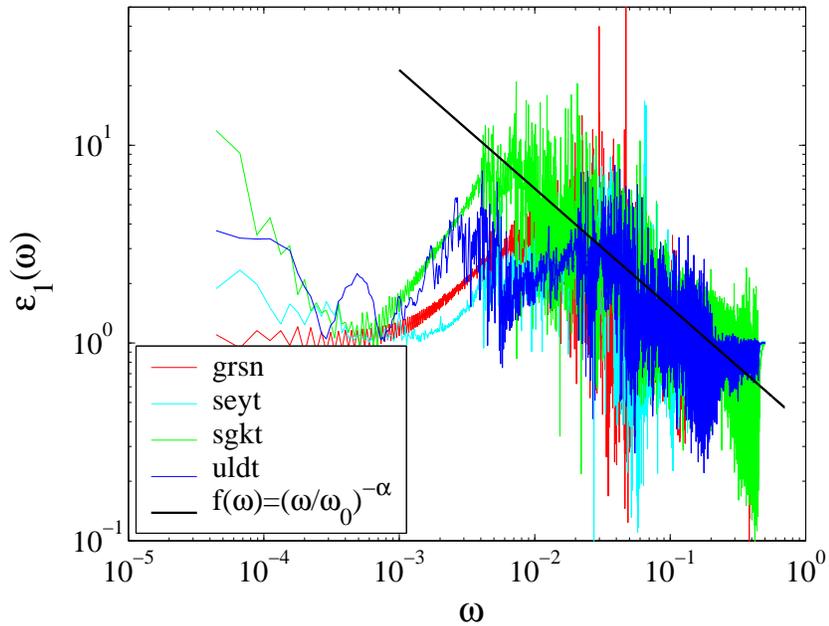}
\end{center}
\caption[kurzform]{\label{Pnm_groupB} (Color online) The frequency dependence of
$\varepsilon_1(\omega)$ for 4 seismic origins: $grsn$, $sgkt$,
$uldt$, $seyt$ in double log-log scale. The time discretization is
$\tau=0.02$ sec. The general power dependence of
$\varepsilon_{1}(\omega)=(\omega/\omega_0)^{-\alpha}$ is submitted
by a continuous line with parameters $\omega_0=0.2$,
$\alpha=0.6$.}
\end{figure}

There is a frequency dependence of $\varepsilon_1(\omega)$ for
4 seismic origins: $grsn$, $sgkt$, $uldt$, $seyt$ in double
log-log scale has been presented in Fig. 4. Discretization time is
$\tau=0.02$ sec. General power  dependence
$\varepsilon_{1}(\omega)=(\omega/\omega_0)^{-\alpha}$ is submitted
by a continuous line with parameters $\omega_0=0.2$, $\alpha=0.6$.

\textbf{Signals of group C} can not be attributed to one of the
above groups. The parameter $\varepsilon_{1}(\omega)$
fluctuate strongly between 1 and 10 in full frequency scale. That
testifies  existence of  strong memory effects in the long-range
time correlation. A possible origin of the similar signals is due
to strong Earthquake. More careful and detailed analysis of the
signal structure on the various time scales and relaxation
levels (with the taking into account of the long-range
correlation, memory effects, nonergodicity and metastability of
underlying system) is required.

\begin{figure}
\begin{center}
\includegraphics[width=0.8\textwidth]{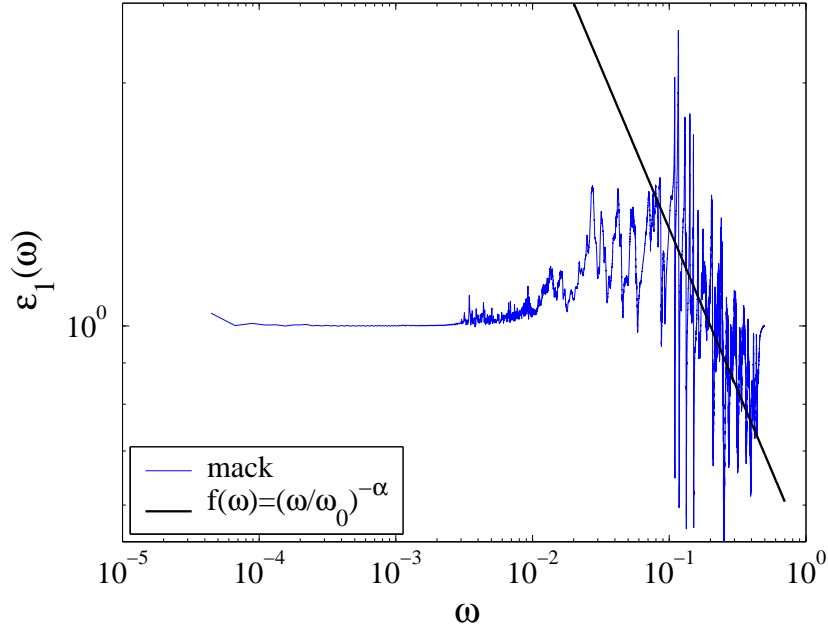}
\end{center}
\caption[kurzform]{\label{Pnm_groupC} (Color online) The frequency dependence of
non-Markovity parameter $\varepsilon_1(\omega)$ for seismic origin
$mack$ in the double log-log scale. The time discretization is $\tau=0.02$ sec.
The general power dependence of
$\varepsilon_{0}(\omega)=(\omega/\omega_0)^{-\alpha}$ is submitted
by a continuous line with parameters $\omega_0=0.2$,
$\alpha=0.4$.}
\end{figure}

Fig. 5 displays the frequency dependence of
$\varepsilon_1(\omega)$ in double log-log scale for $mack$.
Discretization time is $\tau=0.02$ sec. General power dependence
$\varepsilon_{1}(\omega)=(\omega/\omega_0)^{-\alpha}$ is submitted
by a continuous line with parameters $\omega_0=0.2$, $\alpha=0.4$.

The analysis of all spectra shows that all the signals can be
classified into  three different groups in the order of the
breaking of fractal behavior of high frequency dependence of
$\varepsilon_1(\omega)$. Signals for the \textbf{group A} can be
characterized by stronger fractality with the exponents
$\alpha=1.05$. A linear trend with the small fluctuation has been
simultaneously observed in the spectrum. We can see  range of the
diversity $10 < \varepsilon_1(\omega) < 100$. Signals for the
\textbf{group B} are characterized by the breaking of fractality
with the exponents $\alpha=0.6$ . A nonlinear oscillating trend
with big fluctuation has been observed here. Spectra of signals
for \textbf{group C} are characterized by a weak fractality with
the exponent $\alpha=0.4$ and $\varepsilon_1(\omega) \sim 1$.

\begin{figure}
\begin{center}
\includegraphics[width=1.0\textwidth]{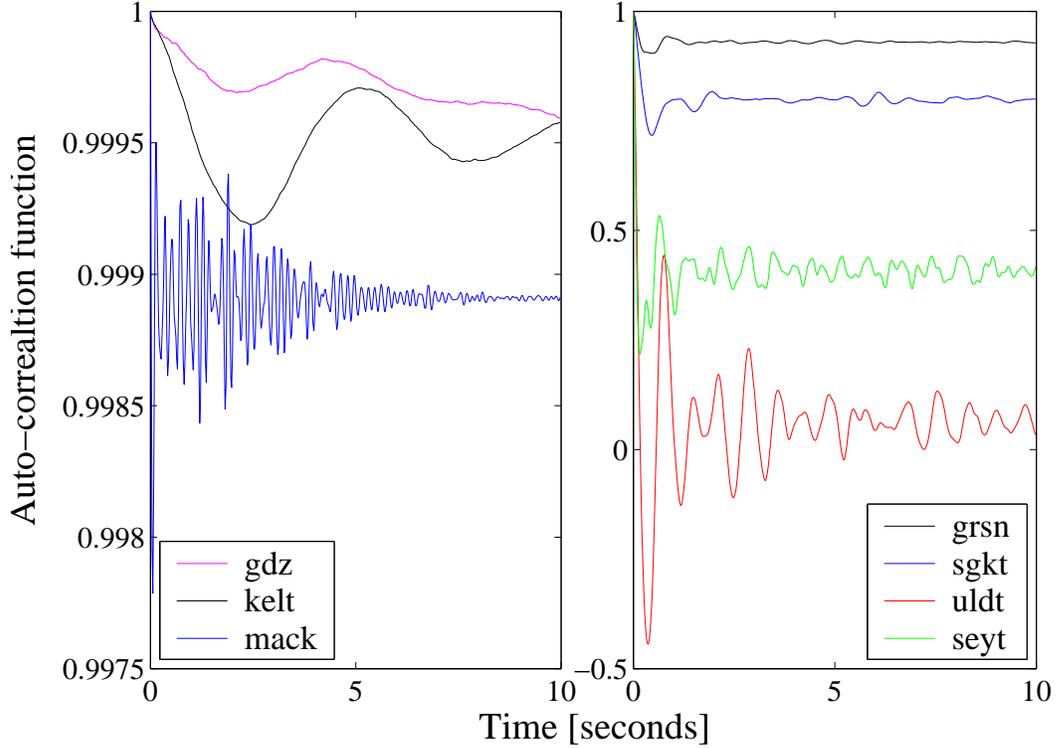}
\end{center}
\caption[kurzform]{\label{ACF} (Color online) The auto-correlation function of the signals.}
\end{figure}

The auto-correlation functions (ACF)
\begin{equation}
C(t)=\frac{\langle \xi(0) \xi(t) \rangle}{\langle |\xi(0)|^2
\rangle}
\end{equation}
for the signals of different groups have been presented in Fig. 6.

The left panel includes ACF for the signals of \textbf{group A}
and \textbf{group C}, ($mack$) and ($kelt$, $gdz$), while the
right panel contains ACF for the signals of \textbf{group B}
($grsn$, $sgkt$, $uldt$, $seyt$). The brackets here note averaging
in time iterations. It is seen from the Figures, that
auto-correlation of the signals has a pronounced nonergodic
character (undamped behavior of the time correlation function at
time $t\rightarrow\infty)$:

\begin{equation}
\lim_{t\rightarrow \infty }C(t)\neq 0.
\end{equation}

According to recent works on the ergodic hypothesis Eqns. (9),
(10) would imply violation of ergodicity. Net results
\cite{Bao} suggest the breaking of ergodicity for a class of
generalized, Brownian motion, obeying a non-Markovian dynamics
being driven by a generalized Langevin equation. This very feature
originates from  vanishing of the effective friction. Khinchin's
theorem of ergodicity is examined \cite{Lee} by means of linear
response theory. The resulting ergodic condition shows that,
contrary to the theorem, irreversibility is not a sufficient
condition for ergodicity.

Similar behavior of the time correlation functions is
characteristic for the supercooled and glass states of condensed
matter \cite{Goetze,Das}. A higher level of nonergodicity
corresponds with the signals of \textbf{group A} and
\textbf{group C}, whereas signals of \textbf{group B} are
characterized by the minor nonergodicity. Signals from the object
$uldt$ are rigorously ergodic. Therefore one can suppose that
these signals cannot belong to Earthquake.


\begin{table}
\begin{center}
\caption{The frequency relaxation parameters for seismic signals.}
\begin{tabular}{l|c|c|c|c|c}
\hline
\hline
 $\textrm{Object}$  &
 $\lambda_1$ &
 $\lambda_2$ &
 $\lambda_3$ &
 $\Lambda_1$ &
 $\Lambda_2$ \\
\hline
grsn & -0.0306 & -0.2004 & -0.8945 & 0.0497  & 0.0396 \\
kelt & -0.0667 & -1.1399 & -1.0485 & -0.0193 & 0.2954 \\
mack & -0.3938 & -0.5808 & -1.0747 & 0.4374 & -0.0342 \\
sgkt & -0.0306 & -0.7490 & -1.0849 & 0.0156 & -0.2967 \\
uldt & -0.0349 & -0.2605 & -0.7609 & 0.0526 & 0.1243 \\
seyt & -0.0745 & -0.2434 & -0.8427 & 0.1173 & 0.0590 \\
gdz  & -0.0646 & -0.8836 & -0.9650 & 0.0155 & 0.2586 \\
\hline
\hline
\end{tabular}
\end{center}
\label{_tab1}
\end{table}

Table 1 presents a set of relaxation parameters
$\lambda_1, \lambda_2 , \lambda_3 , \Lambda_1$ and   $\Lambda_2$
for seismic signals from: $grsn$, $kelt$, $mack$, $sgkt$, $uldt$,
$seyt$, $gdz$. The set of these parameters characterizes some
peculiarities of relaxation processes on the low relaxation levels
of seismic systems (1, 2 and 3). It has seen from the Table that
these parameters don't have a clear distinction vs distinctions
unlike those visible from the frequency dependence of
$\varepsilon_1(\omega)$. Therefore  thinner  and more sensitive
techniques are needed for  studying of dynamic processes in
examined signals.

\begin{figure}
\begin{center}
\includegraphics[width=1.65\textwidth]{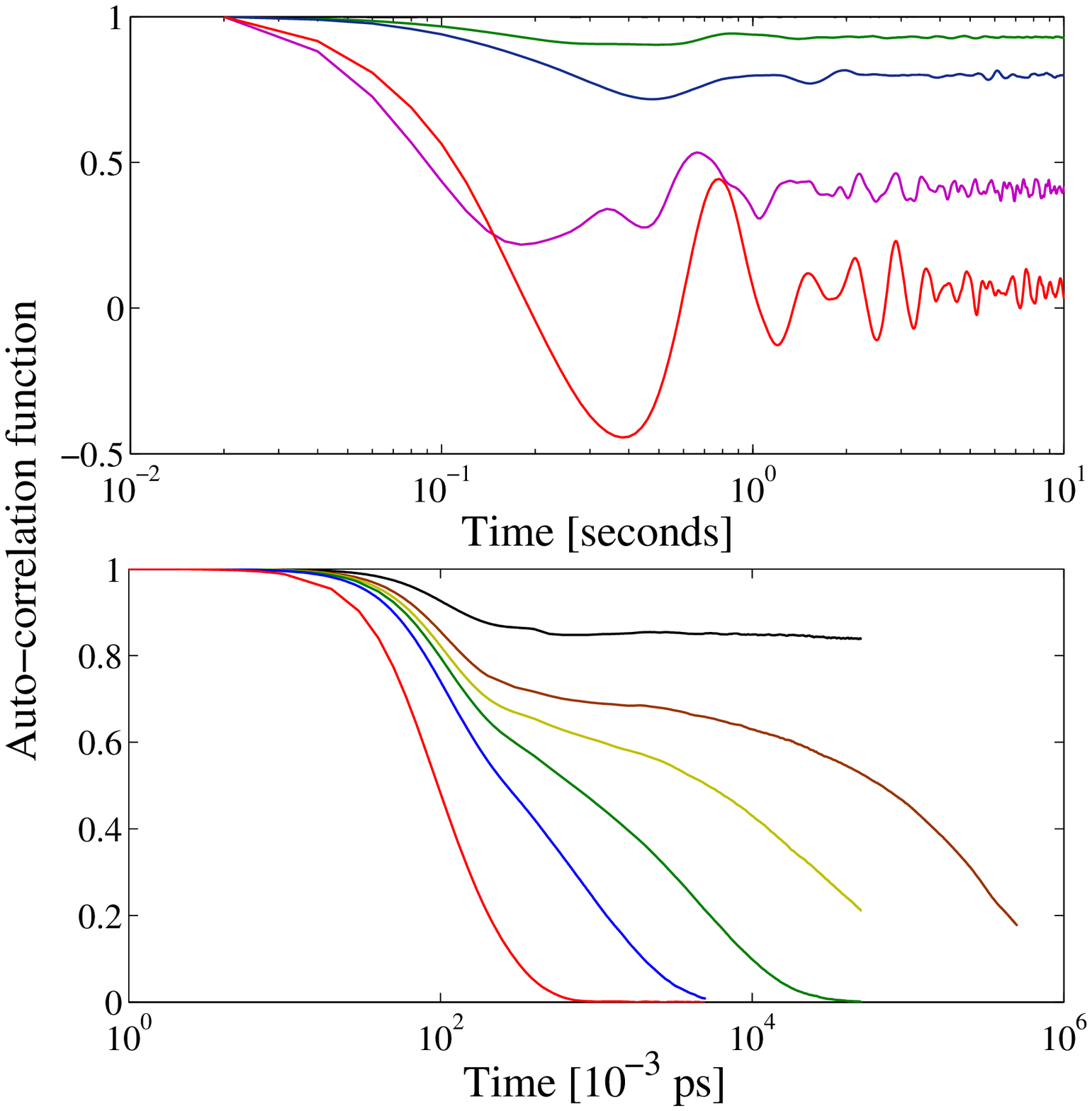}
\end{center}
\caption[kurzform]{\label{ACF} (Color online) On  top, auto-correlation functions of the
seismic events  $grsn$, $sgkt$, $uldt$ and $seyt$ in time
log-scale. At the bottom:  temperature dependence of incoherent
scattering function $F_{s}(k,t)$ for the $Cu$ - component in
metallic glass system  $Al_{50}Cu_{50}$-alloy for the wave vector
$k=3.05\textrm{\AA}^{-1}$ at the temperatures: T=2000 K, 1000 K, 600 K,
500 K, 400 K, 200 K (bottom up).}
\end{figure}

Figure 7 depicts the comparison of seismic data  with results of
computer simulation for the metallic glass $Al_{50}Cu_{50}$. On  top,
auto-correlation functions of the seismic events ($grsn$, $sgkt$,
$uldt$ and $seyt$) are shown. At the bottom,  auto-correlation
functions for incoherent scattering  of copper atoms are shown for
comparison. The data are received with the help of statistical
averaging on ensembles of statistical systems. Each curve  was
received  by  time averaging. Full sample consists of 45000
points. The single calculation was carried out for each separate
time window of 500 points size  with time step of 0.02 s. Further
this window was displaced to one step to the right up to the end
of time sampling.

On comparison data for seismic phenomena with the results of computer simulations
for the glassy system it is visible, that in behaviour of correlation functions
for EQ's nonergodic effects, characteristic for glass-like
behaviour of dense systems are distinctly observed.

To illustrate the general picture of nonergodic singularities in
chaotic seismic systems  the parameter of nonergodicity $f =
\lim\limits_{t\to \infty}c(t) $ for a set of seismic events has
been calculated. Parameter of nonergodicity for a set of seismic
events  appeared equal: 0.9995 ($gdz$), 0.9995 ($kelt$), 0.99885
($mack$),  0.9310 ($grsn$), 0.7241 ($sgkt$), 0.3965 ($uldt$),
0.0603 ($seyt$). The resulted data are evidence of strong
singularity of seismic phenomena in 5 sources  ($gdz$), ($kelt$),
($mack$), ($grsn$), ($sgkt$), they show moderate nonergodicity for
a source ($uldt$) and weak singularity for a source ($seyt$). All
taken together received data  well speak  about wide variety of
effects of nonergodicity in the seismic phenomena. Similar variety
of nonergodicity effects can be very useful and  extremely
effective for the classification of  wide variety  of seismic
phenomena. Note, that the notions of fractality and non-Markovity
have quite extensively been explored in the past, see, for
example, \cite{Turcotte,Barton,Sornette,Sammis}.

\section{Summary}
In this work we have presented the results of statistical analysis
of seismic signals in Turkey for 7 objects  ($grsn$, $kelt$,
$mack$, $sgkt$, $uldt$, $seyt$, $gdz$). Our study was made in the
context of statistical theory on the discrete non-Markov
processes, which is based on the Generalized Langevin Equation
(GLE). It allows to take into account the effects of the
statistical memory, metastability and space-time nonlocality. We
have shown with  the theory that all considered signals can be
divided into  three groups in  order of breaking of the fractal
behavior in high frequency zone of spectrum of non-Markovity
parameter. Signals from A group  ($kelt$, $gdz$) can be
characterized by the pronounced fractality, signals from the B
group  ($grsn$, $sgkt$, $uldt$, $seyt$) can be characterized by
the moderate fractality and signals from the C group   ($mack$)
correspond to weak fractality and powerful non-Markov processes.
From the analysis of the time correlation function we can
confidently certify   hypothesis Abe \cite{Abe} of nonergodic
``glass-like nature'' of seismic signals for Earth activity. On the
other hand aforementioned testifies an information concerning a
wide variety of metastability in seismic phenomena.

\section{Acknowledgments} Here we used seismic data from ``General
Directorate of Disaster Affairs Earthquake Research Department''.
This work was supported by the Grant of RFBR No. 08-02-00123-a (R. Y. and R. Kh.).
The authors also acknowledge the technical assistance of E.R. Nigmatzyanova.


\begin{thebibliography}{10}

\bibitem{Abe} S. Abe, N. Suzuki,
          ArXiv:cond-mat/0305509.

\bibitem{Corral_2006} \'{A}. Corral,
          ArXiv:cond-mat/0604574v1.

\bibitem{Zwanzig} R. Zwanzig, Phys. Rev. {\bf 124}, 1338 (1961).

\bibitem{Mori} H. Mori, Prog. Theor. Phys. {\bf 33}, 423 (1965).

\bibitem{Yulmetyev_2000} R.M. Yulmetyev, P. H\"anggi, and F.
Gafarov, Phys. Rev. E. {\bf 62}, 6178 (2000).

\bibitem{Yulmetyev_2001} R.M. Yulmetyev, F. Gafarov, P. H\"anggi, R. Nigmatullin, and Sh.
Kayumov, Phys. Rev. E. {\bf 64}, 066132 (2001).

\bibitem{Yulmetyev_2005} R.M. Yulmetyev, A. Mokshin, and P.
H\"anggi, Physica A. {\bf 345}, 303 (2005).

\bibitem{Yulmetyev} V. Yu. Shurygin, R. M. Yulmetyev, and V. V. Vorobjev, Phys.
Lett. A {\bf 148}, 199 (1990); V. Yu. Shurygin and R. M. Yulmetyev,
ibid. {\bf 174}, 433 (1993) V. Yu. Shurygin and R. M. Yulmetyev, ZhETF
{\bf 99}, 144 (1991);v. Phys. JETP {\bf 72}, 80 (1991); {\bf 102}, 852 (1992); {\bf 75},
466 (1992).

\bibitem{Varotsos} P. A. Varotsos, N. V. Sarlis, E. S. Skodas, Phys. Rev. E
{\bf 66}, 011902 (2002); {\bf 67}, 021109 (2003).

\bibitem{Fulinski} A. Fulinski, Phys. Rev. E {\bf 58}, 919
(1998); Z. Siwy, A. Fulinski, Phys. Rev. Lett. {\bf 89}, 158101
(2002).

\bibitem{Weron} S. Mercik, K. Weron, Phys. Rev. E {\bf 63},
051910 (2001).

\bibitem{Bao} J. - D. Bao, P. Hanggi, and Y. - Z. Zhuo, Phys. Rev. E {\bf 72},
061107 (2005).
\bibitem{Lee} M. H. Lee, Phys. Rev. Lett. {\bf 98}, 190601 (2007).

\bibitem{Goetze} W. G\"{o}tze, in \textit{Liquids, Freezing, and the Glass Transition},
edited by J.P. Hansen, D. Levesque, and J. Zinn-Justin (North-Holland, Amsterdam, 1991).

\bibitem{Das} S.P. Das, Rev. Mod. Phys. {\bf 76}, 785 (2004).

\bibitem{Turcotte} D. L. Turcotte, Pure. Appl. Geophys. {\bf 131}, 171 (1989).

\bibitem{Barton}  \textit{Fractals in the Earth sciences}, ed. by
C. C. Barton and P. R. La Pointe (Springer, 265 pp, 1995).

\bibitem{Sornette} D. Sornette, A. Sornette, and Chr. Vanneste, in
\textit{Large Scale Structures in Nonlinear Physics, Lecture Note
in Physics, vol. 392, p.p 275-277} (Springer Berlin/ Heidelberg,
1991).

\bibitem{Sammis} C. G. Sammis and D. Sornette, Proc. Nat. Acad. Sci.
{\bf 99}, SUPP1, 2501 (2002).

\end{thebibliography}
\end{document}